\documentclass[iop]{emulateapj}

\shorttitle{The off-centered Seyfert-like emission in the nucleus of NGC 3621}
\shortauthors{Menezes et al.}

\begin{document}

\title{The off-centered Seyfert-like compact emission in the nuclear region of NGC 3621}

\author{R. B. Menezes, J. E. Steiner and Patr\'icia da Silva}

\affil{Instituto de Astronomia Geof\'isica e Ci\^encias Atmosf\'ericas, Universidade de S\~ao Paulo, Rua do Mat\~ao 1226, Cidade Universit\'aria, S\~ao Paulo, SP CEP 05508-090, Brazil;}
\email{robertobm@astro.iag.usp.br}

\begin{abstract}

We analyze an optical data cube of the nuclear region of NGC 3621, taken with the integral field unit of the Gemini Multi-object Spectrograph. We found that the previously detected central line emission in this galaxy actually comes from a blob, located at a projected distance of $2\arcsec\!\!.14 \pm 0\arcsec\!\!.08$ ($70.1 \pm 2.6$ pc) from the stellar nucleus. Only diffuse emission was detected in the rest of the field of view, with a deficit of emission at the position of the stellar nucleus. Diagnostic diagram analysis reveals that the off-centered emitting blob has a Seyfert 2 spectrum. We propose that the line-emitting blob may be a ``fossil'' emission-line region or a light ``echo'' from an active galactic nucleus (AGN), which was significantly brighter in the past. Our estimates indicate that the bolometric luminosity of the AGN must have decreased by a factor of $ \sim 13 - 500$ during the last $\sim 230$ years. A second scenario to explain the morphology of the line-emitting areas in the nuclear region of NGC 3621 involves no decrease of the AGN bolometric luminosity and establishes that the AGN is highly obscured toward the observer but not toward the line-emitting blob. The third scenario proposed here assumes that the off-centered line-emitting blob is a recoiling supermassive black hole, after the coalescence of two black holes. Finally, an additional hypothesis is that the central X-ray source is not an AGN, but an X-ray binary. This idea is consistent with all the scenarios we proposed.

\end{abstract}

\keywords{galaxies: active --- galaxies: nuclei --- galaxies: individual(NGC 3621) --- Techniques: spectroscopic}

\section{Introduction}

NGC 3621 is a late-type galaxy, classified as SA(s)d \citep{vau91}, at a distance of about 6.8 Mpc (NASA Extragalactic Database - NED). \citet{sat07}, using mid-infrared high spectral resolution data obtained with \textit{Spitzer}, discovered an active galactic nucleus (AGN) in this galaxy, revealed by the presence of a centrally concentrated emission of the [Ne V] 14 and 24 $\mu$m lines. The Ne V ion (with an ionization potential of 96 eV) can be generated by the hard AGN continuum but not by the emission from hot stars in H II regions. Therefore, the detection of the [Ne V] emission is considered a proof of an AGN.

\citet{bar09} analyzed \textit{Hubble Space Telescope} (\textit{HST}) images and observed a compact nuclear stellar cluster in NGC 3621. The authors obtained an optical spectrum of this nuclear cluster, using the Echellette Spectrograph and Imager (ESI) at the Keck-II telescope, and verified that its emission-line properties are compatible with those of a Seyfert 2 nucleus. The stellar velocity dispersion of this spectrum is $\sigma = 43 \pm 3$ km s$^{-1}$. Using the Jeans equations, \citet{bar09} also modeled the stellar kinematics of the nuclear cluster and obtained an upper limit for the mass of the central supermassive black hole (SMBH) of $3 \times 10^6 M_{\sun}$.

Using \textit{Chandra} data, \citet{gli09} detected a weak X-ray source, in the 0.5-8 keV band, coincident with the nucleus of NGC 3621. Although the low signal-to-noise ratio (S/N) of the spectrum of this source made it impossible to perform a more detailed analysis, this X-ray observation, together with the optical and mid-infrared observations mentioned before, suggests the presence of a heavily absorbed AGN. \citet{gli09} also detected two other X-ray sources in NGC 3621, located at distances of $\sim 20\arcsec$ from the nucleus, and concluded that these sources are probably associated with intermediate-mass black holes.

In this paper, we report the detection of an off-centered Seyfert-like emitting blob in a data cube of the nuclear region of NGC 3621, obtained with the Gemini Multi-object Spectrograph (GMOS). In Section 2, we describe the observations, the data reduction, and the data treatment. In Section 3, we analyze the data and present the results. In Section 4, we compare our results with those obtained by previous studies and also propose some scenarios to explain the detected off-centered emitting blob. Finally, we draw our conclusions in Section 5.

\begin{figure*}
\epsscale{0.8}
\plotone{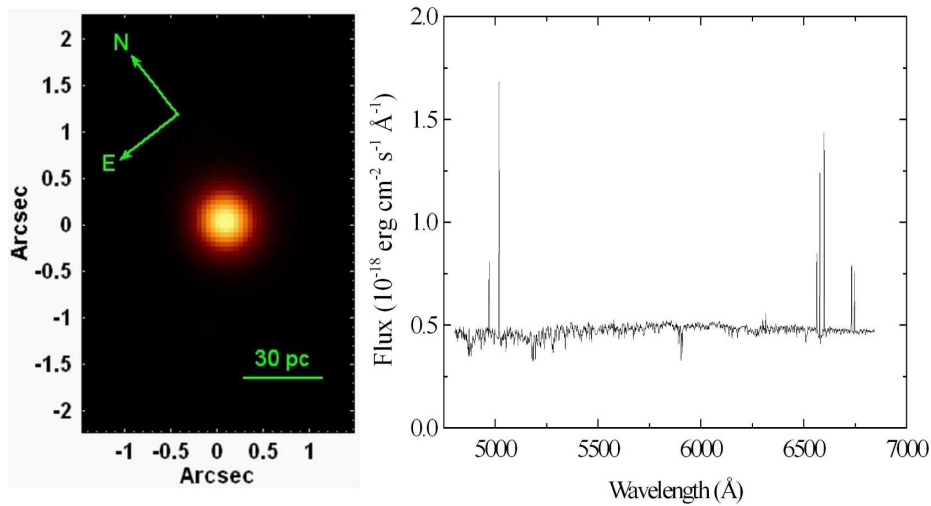}
\caption{Left: image of the treated data cube of NGC 3621, collapsed along the spectral axis. Right: average spectrum of the treated data cube of NGC 3621.\label{fig1}}
\end{figure*}

\section{Observations, reduction and data treatment}

The observations of NGC 3621 were taken on 2014 February 22, using the integral field unit (IFU) of the GMOS at the Gemini-south telescope. We used the IFU in the one-slit mode, which results in a science field of view (FOV) with $5\arcsec \times 3\arcsec\!\!.5$ and in a sky FOV (at a distance of $1\arcmin$ from the science FOV) with $5\arcsec \times 1\arcsec\!\!.75$. Three 815 s exposures, with spatial dithering (with dither steps of $0\arcsec\!\!.2$), were taken using the R831+G5322 grating. The final spectral coverage was $4800 - 6850 \AA$, with a resolution of $R \sim 4300$. The seeing at the night of the observations, based on the acquisition images, was $\sim 0\arcsec\!\!.6$.

Baseline calibration images of GCAL flat, twilight flat, bias, and CuAr lamp were obtained during the observations. The standard star LTT3218 was observed on 2014 March 14. The data were reduced using the Gemini IRAF package. The reduction process included data trim, bias subtraction, cosmic rays rejection (with the L.A.Cosmic routine; van Dokkum 2001), extraction of the spectra, correction for pixel to pixel gain variations (with response curves obtained from the GCAL flat image), correction for fiber to fiber gain variations and for illumination patterns of the instrument (with response maps obtained from the twilight image), wavelength calibration (using the CuAr lamp images), flux calibration (using the spectra of the standard star LTT3218 and taking into account the atmospheric extinction), telluric absorption removal, and data cube construction. At the end of the data reduction, we obtained three data cubes, containing two spatial dimensions and one spectral dimension, with spatial pixels (spaxels) of $0\arcsec\!\!.05$.

We treated the reduced data cubes with a procedure including: correction of the differential atmospheric refraction; calculation of the median of the data cubes; Butterworth spatial filtering \citep{gon02}, to remove high spatial-frequency components from the images of the data cube; instrumental fingerprint removal; and Richardson-Lucy deconvolution \citep{ric72,luc74}, to improve the spatial resolution of the data cube. This entire treatment procedure is described in detail in \citet{men14a,men15}. The FWHM of the point-spread function of the final data cube, estimated from a comparison with a convolved \textit{V} band image obtained with the Wide-field Planetary Camera 2 of the \textit{HST}, is $\sim 0\arcsec\!\!.4$. An image of the treated data cube collapsed along the spectral axis, together with the average spectrum of the data cube, is shown in Figure~\ref{fig1}. The average spectrum shows prominent emission lines. The bright compact area in the image of the collapsed data cube corresponds to the nuclear stellar cluster, previously analyzed by \citet{bar09}.

\begin{figure}
\epsscale{1.20}
\plotone{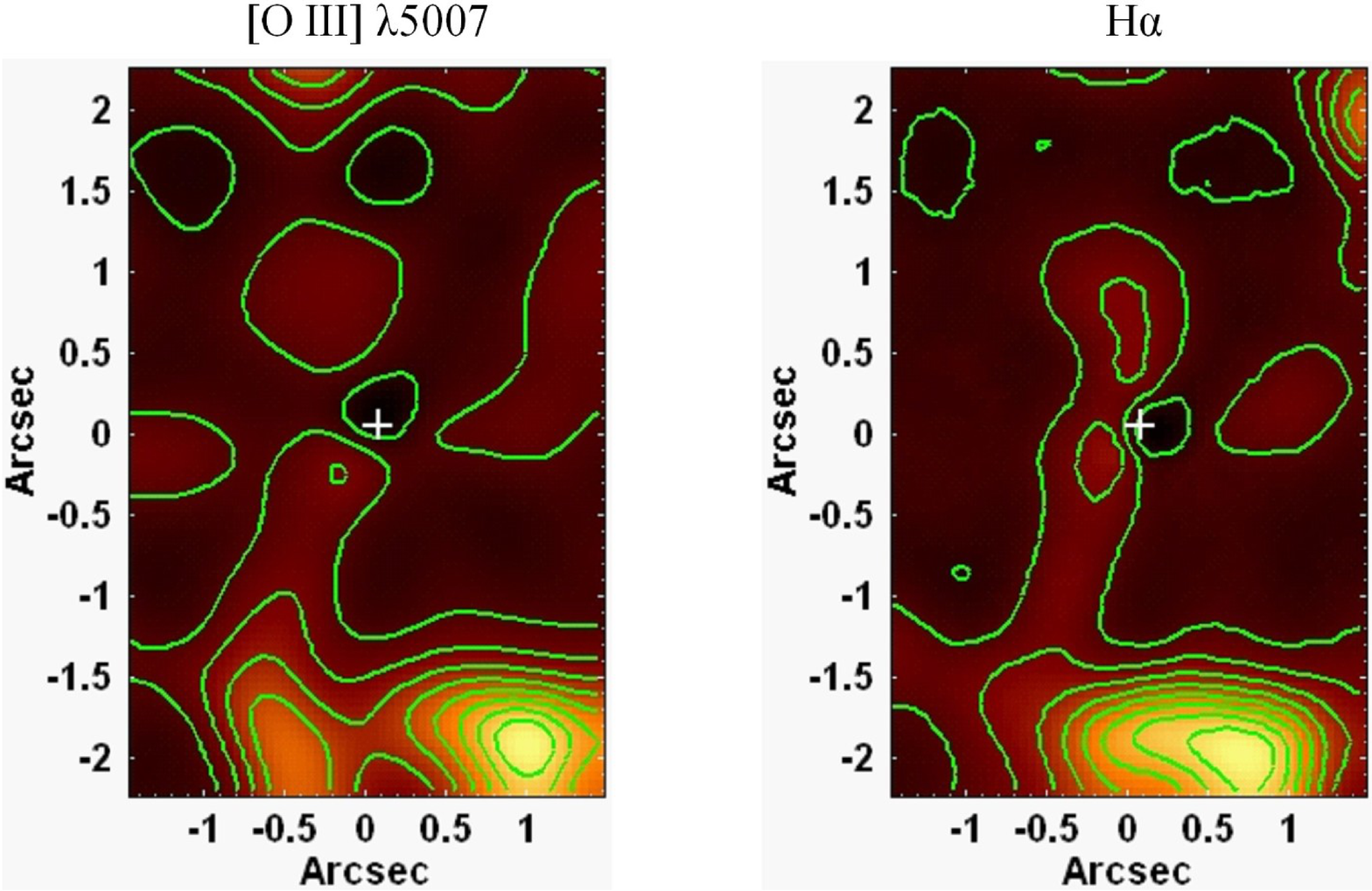}

\vspace{3 mm}

\plotone{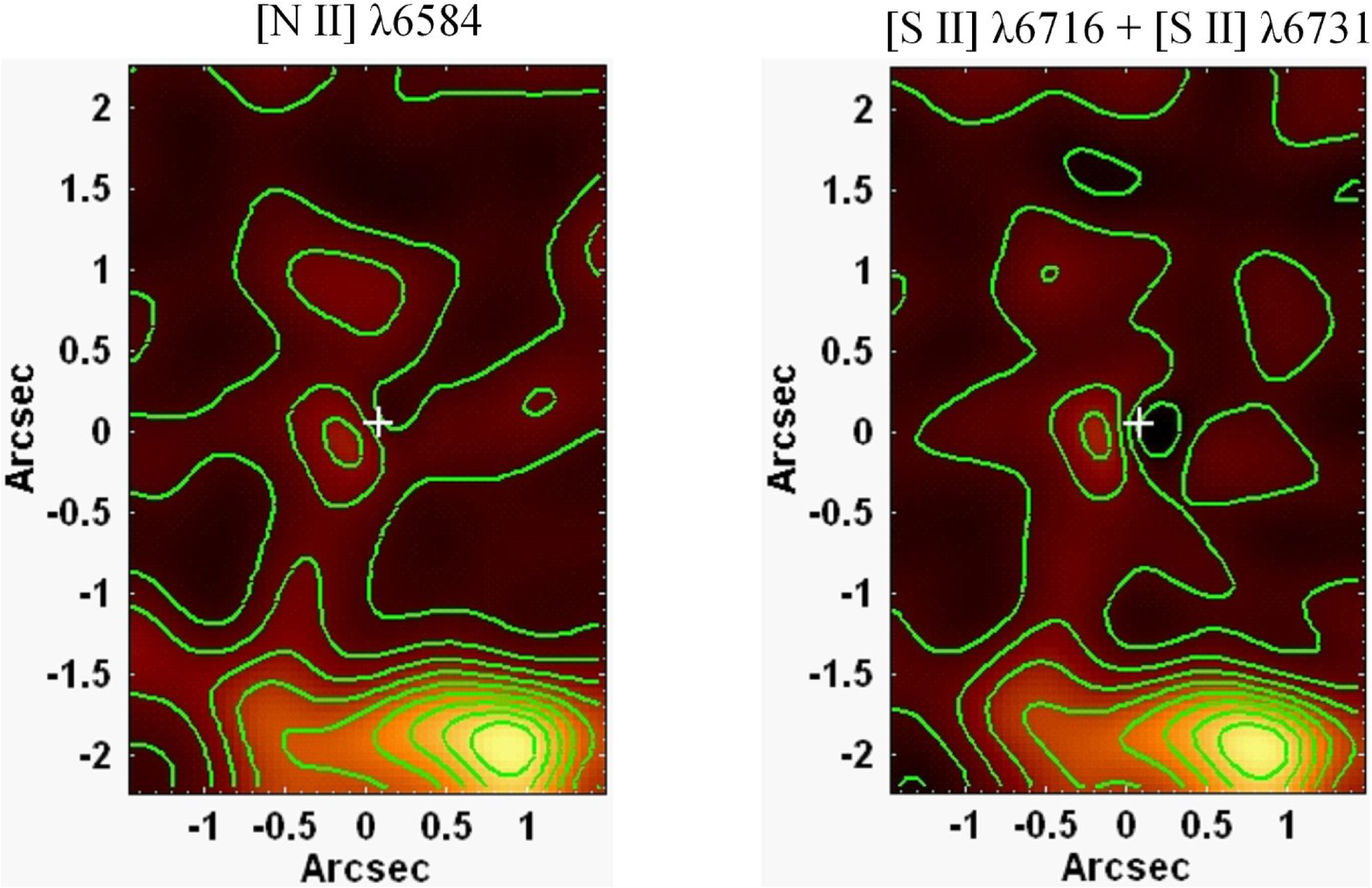}
\\

\vspace{3 mm}

\epsscale{0.61}
\plotone{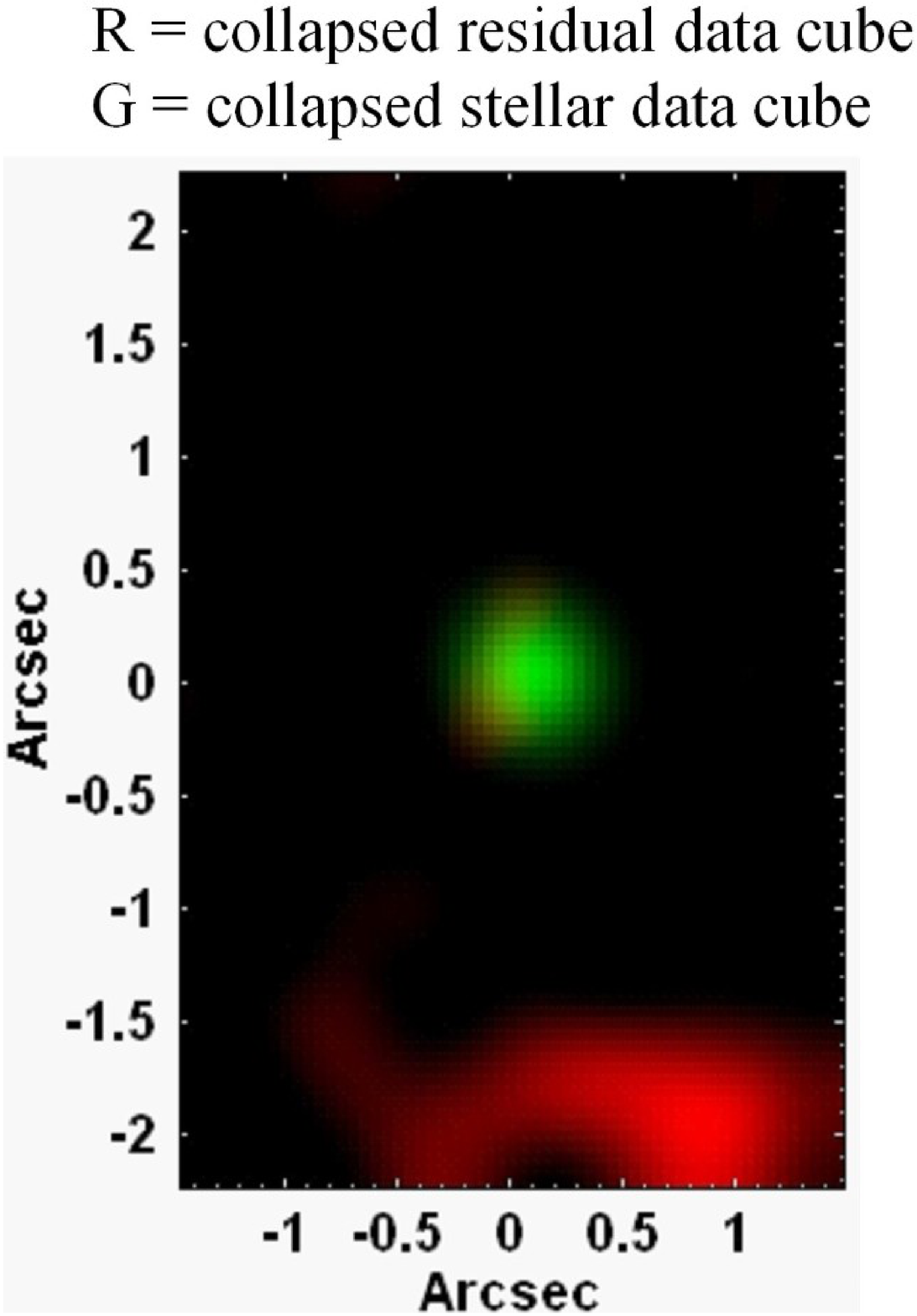}
\caption{Images of the main emission lines of the residual data cube of NGC 3621, obtained after the starlight subtraction, with isocontours and the position of the stellar nucleus marked with a cross. The image at the bottom is an RG composite, with the collapsed stellar data cube (constructed with the synthetic spectra provided by the spectral synthesis) shown in green and the collapsed residual data cube shown in red.\label{fig2}}
\end{figure}

\section{Data analysis and results}

\begin{figure*}
\epsscale{1.1}
\plotone{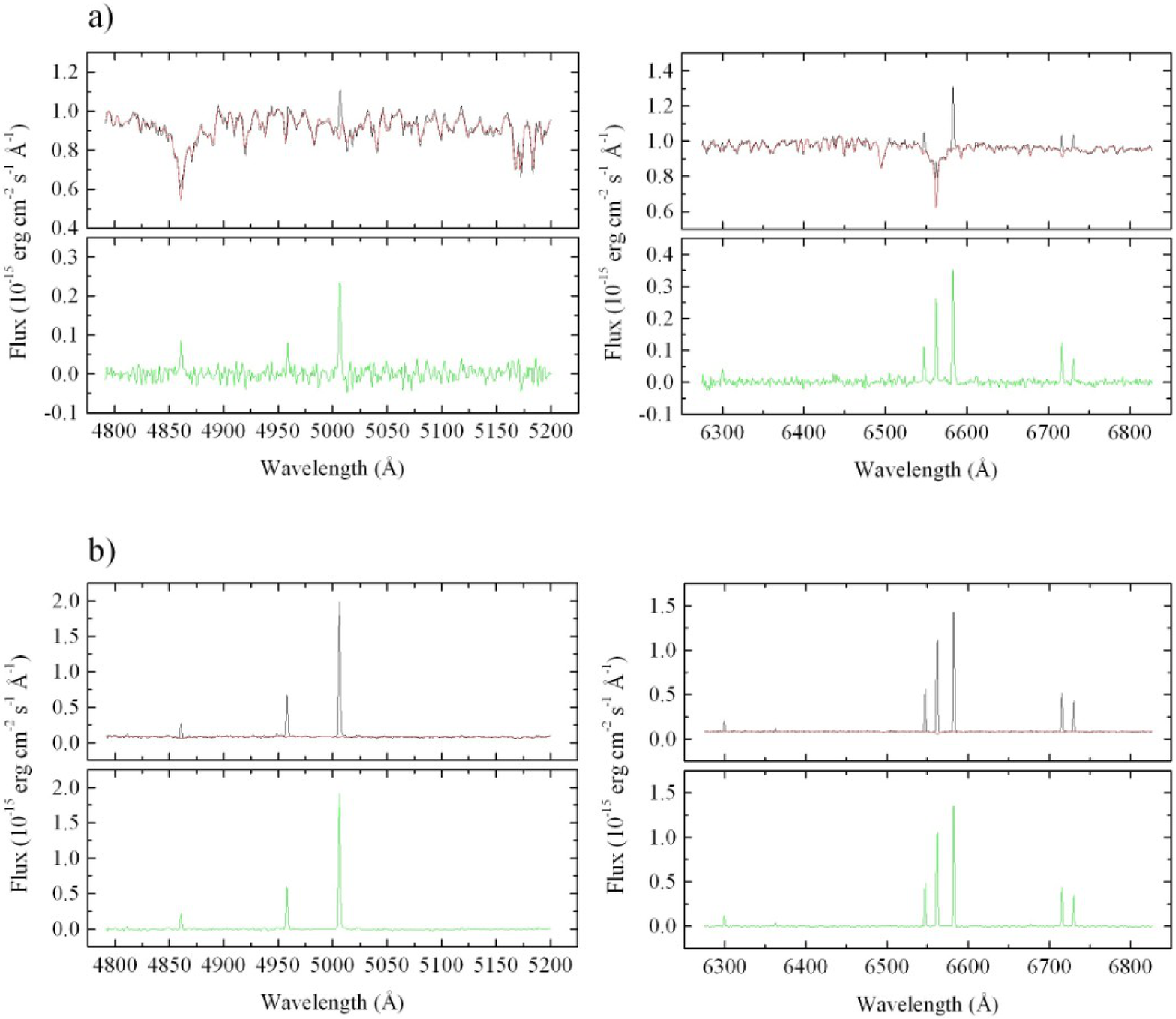}
\caption{Different wavelength ranges of spectra extracted from two circular regions, with radii of $0\arcsec\!\!.5$, centered (a) on the nuclear stellar cluster and (b) on the off-centered emitting blob of NGC 3621. The fits provided by the spectral synthesis are shown in red and the fit residuals are shown in green. The fit residuals in (a) and (b) reveal the nuclear diffuse emission and the emission-line spectrum of the off-centered emitting blob, respectively.\label{fig3}}
\end{figure*}

A detailed analysis of the emission lines in the data cube of NGC 3621 requires an accurate starlight subtraction. In order to do that, we used the Starlight software \citep{cid05} to apply a spectral synthesis to the spectrum of each spaxel of the data cube. This software fits a stellar spectrum with a combination of template spectra from a base. For this work, we used a base of stellar population spectra based on Medium-resolution Isaac Newton Telescope Library of Empirical Spectra (MILES; S\'anchez-Bl\'azquez et al. 2006). We chose the MILES library because of the similarity between its spectral resolution (FWHM = $2.3\AA$) and our spectral resolution (FWHM = $1.3\AA$). The base used for the spectral synthesis contains spectra of 150 stellar populations, with ages between $1.0 \times 10^6$ years and $1.3 \times 10^{10}$ years and with metallicities between 0.0001 and 0.05 (the solar metallicty being $Z_{\sun} = 0.02$). The Starlight software provides the values of: the stellar radial velocity ($V_*$), the stellar velocity dispersion ($\sigma_*$), the extinction at the observed object ($A_V$), and the flux fraction corresponding to each stellar population taken into account in the fit. The spectral synthesis also results in a synthetic stellar spectrum for each observed spectrum.

Before the spectral synthesis was applied to the data cube of NGC 3621, we performed a correction of the Galactic extinction, using $A_V = 0.221$ mag (NED) and the extinction law of \citet{car89}. The spectra were also shifted to the rest frame, using $z = 0.002435$ (NED), and re-sampled with $\Delta\lambda = 1\AA$ per spectral pixel. We then applied the spectral synthesis to the data cube and subtracted the obtained synthetic stellar spectra from the observed ones, which resulted in a data cube with only emission lines. 

In the residual data cube, most of the line emission comes from a compact area southeast from the nucleus. The rest of the FOV shows a weak diffuse emission. The projected distance between the stellar nucleus and the line-emitting blob, calculated using the image of the collapsed residual data cube, is $2\arcsec\!\!.14 \pm 0\arcsec\!\!.08$ (which is equivalent to $70.1 \pm 2.6$ pc), at a position angle (PA) of $163\degr \pm 4\degr$. Figure~\ref{fig2} shows images of the strongest emission lines, with isocontours. All these images have similar morphologies and the isocontours reveal clearly the presence of the blob southeast from the nucleus. There is also a significant extended emission northeast from blob. One important characteristic that can be seen in the images in Figure~\ref{fig2} is that there is a deficit of line emission at the position of the nuclear stellar cluster. There are variations in the exact position of the line-emitting blob for the different lines, which were taken into account to estimate the uncertainty for the projected distance between the nucleus and the blob. Figure~\ref{fig2} also contains an RG composite image, with the collapsed stellar data cube (constructed with the synthetic spectra provided by the spectral synthesis) shown in green and the collapsed residual data cube shown in red.

\begin{figure*}
\epsscale{1.0}
\plotone{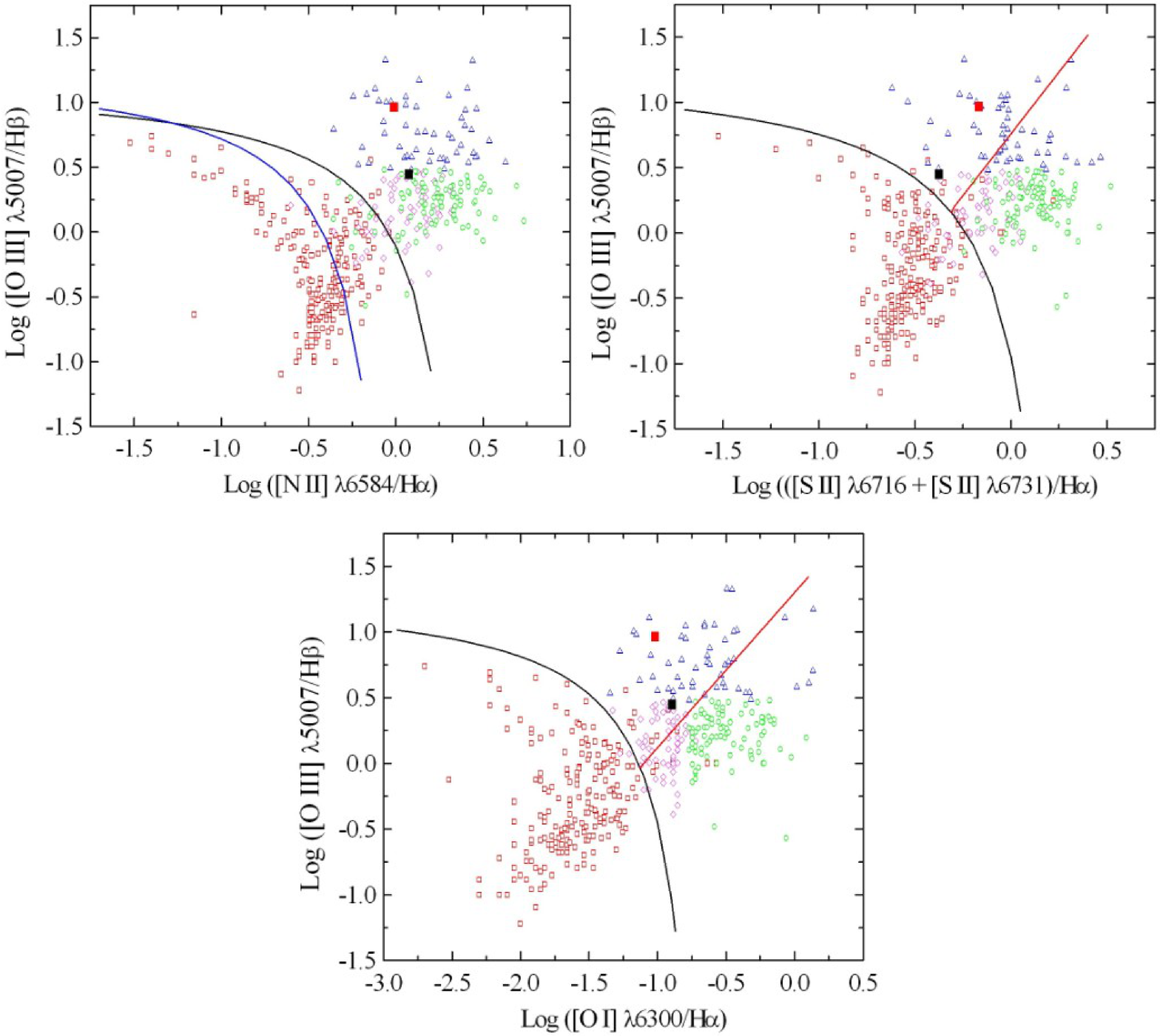}
\caption{Diagnostic diagrams showing the points corresponding to the spectra extracted from the nucleus (filled black square) and from the off-centered emitting blob (filled red square) of NGC 3621. The error bars are comparable to the sizes of the points. The other points correspond to objects analyzed by \citet{ho97}. The open red squares represent H II regions, the open green circles correspond to low-ionization nuclear emission-line regions (LINERs), the open blue triangles represent Seyfert galaxies, and the open magenta diamonds correspond to transition objects. The black curve is the maximum limit for the ionization by a starburst, obtained by \citet{kew01}, the blue curve is the empirical division between H II regions and AGNs, proposed by \citet{kau03}, and the red curve is the division between LINERs and Seyfert galaxies, proposed by \citet{kew06}.\label{fig4}}
\end{figure*}

We extracted the spectra from two circular areas of the data cube of NGC 3621, before the starlight subtraction, with radii of $0\arcsec\!\!.5$. One of these areas was centered on the nuclear stellar cluster and the other was centered on the line-emitting blob. The extracted spectra, together with the fits obtained with the spectral synthesis and the fit residuals, are shown in Figure~\ref{fig3}. One can see that the spectral fits have a very good quality and provided a reliable starlight subtraction. No broad component of the H$\alpha$ and H$\beta$ emission lines was detected. We corrected the extracted spectra for the interstellar extinction at the observed galaxy, using the $A_V$ values provided by the spectral synthesis. After that, using the extinction corrected fit residuals, we calculated emission-line ratios for the nuclear diffuse emission and for the line-emitting blob. The results are shown in Table~\ref{tbl1}. Using the calculated emission-line ratios, we constructed diagnostic diagrams \citep{bal81}, shown in Figure~\ref{fig4}, also including the points corresponding to the objects analyzed by \citet{ho97}. The point representing the blob falls in the branch of the Seyfert galaxies. Therefore, we conclude that the spectrum of this off-centered region is characteristic of a Seyfert 2 and, as a consequence, the emission from this area is related to the presence of an AGN in NGC 3621. The luminosity and the FWHM (corrected for the instrumental spectral resolution) of the [O III] $\lambda 5007$ emission line of this off-centered Seyfert-like emitting blob are $L_{[O III]} = (1.79 \pm 0.09) \times 10^{38}$ erg s$^{-1}$ and FWHM$_{[O III]} = 81 \pm 5$ km s$^{-1}$, respectively. The classification of the spectrum containing the nuclear diffuse emission requires more attention. Following the criteria of \citet{kew06}, we verify that this spectrum is also characteristic of a Seyfert galaxy, but with a lower ionization degree, indicating that the AGN emission in this area has a lower ionization parameter \citep{fer83, hal83}. We can also see that many galaxies classified by \citet{ho97} as transition objects fall very close to the point corresponding to the nuclear spectrum in the diagnostic diagrams. This suggests that the nuclear spectrum may be contamined by the emission from H II regions. It is important to mention that the Balmer decrements (i.e. H$\alpha$/H$\beta$) of the extracted spectra, after the correction for the interstellar extinction at the galaxy, are compatible, at the 1$\sigma$ level, with 2.86, which corresponds to Case B recombination, with a temperature of $10^4$ K and an electron density of $10^2$ cm$^{-3}$ \citep{ost06}.

\begin{table*}
\begin{center}
\caption{Emission-line ratios obtained for the spectra shown in Figure~\ref{fig3}.\label{tbl1}}
\begin{tabular}{ccc}
\tableline\tableline
Ratio & Nuclear Diffuse Emission & Off-centered Emitting blob \\
\tableline
$[$N II$]$ $\lambda 6584$/H$\alpha$ & $1.19 \pm 0.09$ & $0.97 \pm 0.07$ \\
$[$O III$]$ $\lambda 5007$/H$\beta$ & $2.9 \pm 0.7$ & $8.5 \pm 0.6$ \\
($[$S II$]$ $\lambda 6716 + \lambda 6731$)/H$\alpha$ & $0.66 \pm 0.04$ & $0.65 \pm 0.05$ \\
$[$O I$]$ $\lambda 6300$/H$\alpha$ & $0.13 \pm 0.03$ & $0.104 \pm 0.014$ \\
H$\alpha$/H$\beta$ & $2.5 \pm 0.6$ & $2.92 \pm 0.21$ \\
$[$S II$]$ $\lambda6716$/$[$S II$]$ $\lambda 6731$ & $1.27 \pm 0.09$ & $1.20 \pm 0.09$ \\
\tableline
\end{tabular}
\end{center}
\end{table*}

In order to perform an analysis of the stellar kinematics around the stellar nucleus and around the off-centered emitting blob, we used the penalized pixel fitting (pPXF) procedure \citep{cap04}, which, similarly to the spectral synthesis, also fits the stellar spectrum of an object with a combination of template spectra. However, in this case, the template spectra are convolved with a Gauss-Hermite expansion. The values of the stellar radial velocity ($V_*$), the stellar velocity dispersion ($\sigma_*$), and the Gauss-Hermite coefficients $h_3$ and $h_4$ of the observed spectrum are provided by the pPXF method. We used, again, a base of stellar population spectra based on MILES. However, before this procedure was applied, first of all, we extracted a spectrum from a circular area of the data cube, before the starlight subtraction, centered on the emitting blob, with a radius of $0\arcsec\!\!.2$. Then, the spectrum from an adjacent circular area (with the same radius) was also extracted. The result obtained by subtracting this second spectrum from the first one did not reveal clear traces of absorption lines. This indicates that the emitting blob does not show distinct spectral stellar properties in comparison to the surroundings. Therefore, the value of $V_*$ in this area is a property of the host galaxy and not of the emitting region. We verified that the pPXF method only provides sufficiently reliable values of $V_*$ and $\sigma_*$ for stellar spectra with S/N of, at least, 10. However, the S/N (in the wavelength range $5590\AA - 5740\AA$) of the spectrum of the emitting blob extracted from a circular area with a radius of $0\arcsec\!\!.2$ is $\sim 5$. A minimum extraction radius of $0\arcsec\!\!.5$ is required to obtain a S/N of 10. Since the spectra of the stellar nucleus and of the emitting blob were extracted from circular regions with this same radius, we applied the pPXF to these two spectra. We verified that the stellar radial velocity of the emitting blob, relative to the nucleus, is $V_*(blob) = -10 \pm 3$ km s$^{-1}$. In adition, the stellar velocity dispersion values at the positions of the stellar nucleus and of the emitting blob are $\sigma_*(nucleus) = 50 \pm 3$ km s$^{-1}$ and $\sigma_*(blob) = 47 \pm 11$ km s$^{-1}$, respectively. One should note that this value of $\sigma_*(blob)$ is compatible with the one obtained by \citet{bar09}, at the 2$\sigma$ level. We also fitted a Gaussian function to the [N II] $\lambda 6584$ emission line, after the starlight subtraction, and obtained a gas radial velocity of the blob, relative to the nucleus, of $V_{gas}(blob) = -46 \pm 5$ km s$^{-1}$. This value is not compatible, even at the 3$\sigma$ level, with $V_*(blob)$, indicating that the gas kinematics in the emitting blob is apparently disconnected from the stellar kinematics in this area. It is important to mention that we did not construct maps of the gas radial velocity and of $V_*$ because the S/N ratios of the stellar continuum and amplitude/noise (A/N) ratios of the main emission lines were significantly small in many areas of the FOV. We tried to apply a spatial binning to increase these S/N and A/N ratios; however, this procedure reduced considerably the spatial resolution of the data cube and the resulting kinematic maps were not worthy of publication.

\begin{figure*}
\epsscale{1.0}
\plotone{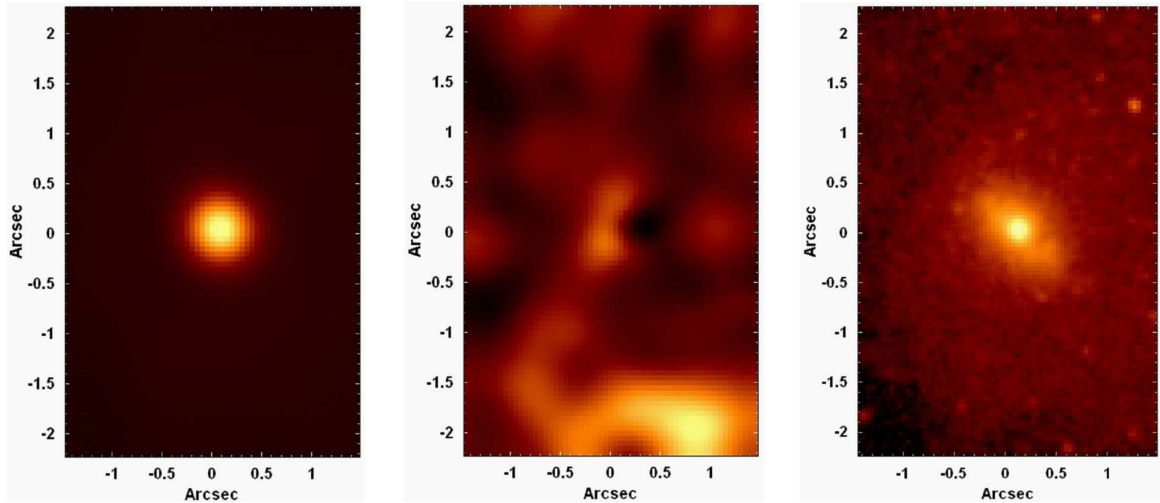}
\caption{Left: image of the data cube of NGC 3621, before the starlight subtraction, collapsed along the spectral axis (the same image shown in Figure~\ref{fig1}). Middle: image of the data cube of NGC 3621, after the starlight subtraction, collapsed along the spectral axis (the same image shown in red in the RG composition in Figure~\ref{fig2}). Right: image of the nuclear region of NGC 3621 (with the same FOV of the GMOS-IFU data cube), obtained with WFC3 of the \textit{HST}, in the F336W filter (\textit{U} band).\label{fig5}}
\end{figure*}

The spectral synthesis of the data cube of NGC 3621 revealed that the stellar emission, both from the nuclear stellar cluster and from the surroundings, is due to stars formed in two episodes: a low metallicity ($Z = 0.004 - 0.008$) star formation with $T = (1.3 - 2.5) \times 10^9$ years and a high metallicity star formation ($Z = 0.02 - 0.05$) with $T = (5 - 9) \times 10^8$ years.

In order to look for the presence of hot stars in the nuclear region of NGC 3621, we retrieved an image of this galaxy, obtained with the Wide-field Camera 3 (WFC3) of the \textit{HST}, in the F336W filter (\textit{U} band) \citep{car15}. Figure~\ref{fig5} shows a comparison between the images of the data cubes of NGC 3621, before and after the starlight subtraction, collapsed along their spectral axis and the WFC3 image (with the same FOV of the data cube). The line-emitting blob is not visible in the WFC3 image, which shows structures, along the FOV, probably associated with hot stars. Some of these structures are located very close to the nucleus and may be responsible for, at least, part of the ionization of the gas in that area. This is compatible with the fact that the diagnostic diagram analysis for the nuclear diffuse emission suggests a contamination by the emission from H II regions. On the other hand, the WFC3 image reveals no stellar concentration in the area of the line-emitting blob. Therefore, we conclude that the emission from hot stars is probably not significant for the ionization of the gas in the emitting blob, which is also compatible with the results obtained with the diagnostic diagram analysis.

\section{Discussion and comparison with previous studies}

The deficit of line emission at the position of the nucleus of NGC 3621, together with the detection of an off-centered line-emitting blob (with a Seyfert 2 spectrum), at a projected distance of $2\arcsec\!\!.14 \pm 0\arcsec\!\!.08$ from the nucleus, has important implications for the studies of this galaxy. As mentioned before, the discovery of an AGN in NGC 3621 was made by \citet{sat07}, using \textit{Spitzer} data. However, these authors did not have enough spatial resolution to conclude that the line emission they detected was indeed coming from the nucleus. The [O III] $\lambda 5007$/H$\beta$ and [N II] $\lambda 6584$/H$\alpha$ emission-line ratios obtained by \citet{bar09}, from a spectrum observed with a $0\arcsec\!\!.75$-wide slit of the ESI at the Keck-II telescope, are compatible with the values calculated by us for the off-centered emitting blob, at the 1$\sigma$ level. In addition, the ([S II] $\lambda 6716 + \lambda 6731$)/H$\alpha$ ratio obtained by \citet{bar09} is also compatible with our value, at the 2$\sigma$ level. As \citet{bar09} extracted their spectrum with an aperture of $5\arcsec$ and with a PA of the slit of $9\degr$, the area covered by the slit during the observations included the off-centered emitting blob. Therefore, we conclude that the emission detected by \citet{sat07} and by \citet{bar09} was probably originated in the off-centered line-emitting blob and not in the nucleus of NGC 3621. Similarly to what we observed here, the \textit{HST} images of this galaxy analyzed by \citet{bar09} did not reveal traces of the off-centered emitting blob. This indicates that the blob is not a significant continuum emiter or it is highly obscured.

The central X-ray source detected by \citet{gli09} (R.A. = $11^h 18^m 16^s\!\!.51$, decl. = $-32\degr 48\arcmin 50\arcsec\!\!.4$) is located at projected distances of $0\arcsec\!\!.25$ and $0\arcsec\!\!.41$ from the positions of the nucleus of NGC 3621 obtained from the 2MASS catalog (R.A. = $11^h 18^m 16^s\!\!.50$, decl. = $-32\degr 48\arcmin 50\arcsec\!\!.6$; Skrutskie et al. 2006) and from the WFC3 image shown in Figure~\ref{fig5} (R.A. = $11^h 18^m 16^s\!\!.525$, decl. = $-32\degr 48\arcmin 50\arcsec\!\!.78$, with an uncertainty of $0\arcsec\!\!.04$), respectively. Although \citet{gli09} did not provide error bars for the position of the X-ray source, we believe that a positional accuracy of $\sim 0\arcsec\!\!.3$ should be possible (considering that the authors detected $\sim 20$ counts in that source). So, if we assume an uncertainty of $0\arcsec\!\!.3$ for the position of the central X-ray source, such position is compatible with the nuclei obtained from the 2MASS catalog and from the WFC3 image, at 1$\sigma$ and 2$\sigma$ levels, respectively. This suggests the presence of an AGN in the nucleus of NGC 3621. Considering that, the first scenario we propose to explain the morphology of the emitting areas in the nuclear region of NGC 3621 assumes that the line-emitting blob is part of the narrow-line region (NLR) of the AGN, which is located at the stellar nucleus. Using the image of the collapsed data cube of NGC 3621, after the starlight subtraction, we estimated the projected opening angle of the ionization cone needed to generate the entire off-centered emission-line structure and obtained a value of $67\degr \pm 11\degr$. The deficit of emission at the nucleus could be the result of a recent decrease of the luminosity of the AGN. In that case, the observed line-emitting blob can be interpreted as a ``fossil'' emission-line region or a light ``echo'' from the previously brighter AGN in the galaxy. In order to evaluate this hypothesis, we tried to reproduce the observed emission-line ratios of the off-centered blob with a simulation performed with version 13.03 of the Cloudy software, last described by \citet{fer13}. In this simulation, we assumed a central AGN, with a power law spectrum with a spectral index of $\alpha = -1.5$. We also assumed that the radius between the emitting region and the central AGN is equal to the projected distance between the stellar nucleus and the emitting blob ($70.1$ pc) observed in the data cube of NGC 3621. The density of the emitting region was taken as being equal to $240$ cm$^{-3}$, which was obtained from the [S II] lines of the emitting blob spectrum, assuming a temperature of 10,000 K. Finally, a filling factor of 0.01 was considered. The bolometric luminosity of the central AGN and the metallicity of the gas were varied and the simulations were repeated, in order to reproduce the observed emission-line ratios. At the end, considering a bolometric luminosity for the AGN of $L_{bol}(simulated) = 4 \times 10^{41}$ erg s$^{-1}$ and a metallicity of $Z = 1.4 Z_{\sun}$, we obtained [O III] $\lambda 5007$/H$\beta = 9.00$, $([S II] \lambda 6717 + \lambda 6731$)/H$\alpha = 0.70$, [N II] $\lambda 6584$/H$\alpha = 0.88$, and [O I] $\lambda 6300$/H$\alpha = 0.12$. The [O III] $\lambda 5007$/H$\beta$, $([S II] \lambda 6717 + \lambda 6731$)/H$\alpha$ and [O I] $\lambda 6300$/H$\alpha$ values are compatible with the observed ratios at the 1$\sigma$ level, while the [N II] $\lambda 6584$/H$\alpha$ value is compatible with the observed ratio at the 2$\sigma$ level. Our simulation also resulted in the emission of the [Ne V] 14 and 24 $\mu$m lines, which is consistent with the observations made by \citet{sat07}. Again, assuming that the radius between the line-emitting blob and the AGN is equal to the projected distance between the blob and the stellar nucleus, we conclude that any decrease in the AGN luminosity must have occurred during the last $\sim 230$ years. Using the relation $L_{bol} < L_{Edd}$ (where $L_{Edd}$ is the Eddington luminosity), the value provided by our simulation for $L_{bol}(simulated)$ implies that the mass of the central supermassive black hole (SMBH) in NGC 3621 must satisfy $M_{bh} > 3000 M_{\sun}$, which is compatible with the estimate of $M_{bh} < 3 \times 10^6 M_{\sun}$ obtained by \citet{bar09}.

The low count rate of the central X-ray source detected by \citet{gli09} prevented the authors to perform a detailed spectral characterization of the source. Nevertheless, using simulation tools, the authors verified that, assuming column densities of $N_H(low) = 5 \times 10^{21}$ cm$^{-2}$ (consistent with the value derived from the spectral fitting of the two off-nuclear X-ray sources in this galaxy) and $N_H(high) = 5 \times 10^{23}$ cm$^{-2}$ (typical of non Compton-thick Seyfert 2 galaxies; Bassani et al. 1999), the intrinsic X-ray luminosities of the source are $L_{2-10 keV}(low) = 5 \times 10^{37}$ erg s$^{-1}$ and $L_{2-10 keV}(high) = 2 \times 10^{39}$ erg s$^{-1}$, respectively. Using the relation $L_{bol} = 16 \times L_{2-10 keV}$ \citep{ho08}, we obtain current bolometric luminosities of $L_{bol}(low) = 8 \times 10^{38}$ erg s$^{-1}$ and $L_{bol}(high) = 3.2 \times 10^{40}$ erg s$^{-1}$, in the moderate and heavy absorption regimes, respectively, for the central AGN in NGC 3621. Therefore, in our AGN shutdown scenario, considering a moderate absorption regime, the AGN bolometric luminosity must have decreased by a factor of $\sim 500$, during the last $\sim 230$ years. On the other hand, considering a heavy absorption regime, the AGN bolometric luminosity must have decreased by a factor of $\sim 13$, during the last $\sim 230$ years.

A problem of the AGN shutdown scenario is that a situation in which a compact X-ray source is detected at the position of the AGN but only diffuse emission is detected in the optical is very unusual. A possible explanation to that is a considerable obscuration of the central source, which reduces drastically the emission in the optical toward the observer (however, the emission toward the emitting blob is not significantly attenuated). This hypothesis is consistent with the heavy absorption regime described above. The difference between the gas radial velocity ($V_{gas}(blob) = -46 \pm 5$ km s$^{-1}$) and the stellar radial velocity ($V_*(blob) = -10 \pm 3$ km s$^{-1}$) at the position of the line-emitting blob suggests that the blob may be part of an outflow caused by the AGN. Another possiblity is that the gas of the emitting blob lies out of the plane of the galaxy's disk. Considering that the bolometric luminosity decrease in the heavy absorption regime is not so dramatic, a second scenario to explain the morphology of the line-emitting areas in the data cube of NGC 3621 involves no decrease of the AGN bolometric luminosity. In that case, the AGN must be highly obscured toward the observer but not toward the line-emitting blob.

Scenarios very similar to the ones mentioned above, but on a much larger scale, have been proposed to explain the detected emission from the irregular gas cloud SDSS J094103.80+344334.2, also known as ``Hanny's Voorwerp''. This object is located at a distance of $\sim 25$ kpc, toward southeast, from the galaxy IC 2497 \citep {joz09, lin09}. The line emission from this gas cloud (characteristic of a Seyfert 2 galaxy), together with the lack of X-ray emission from the nucleus of IC 2497, has lead to the hypothesis that Hanny's Voorwerp is a light echo from the AGN in IC 2497, which has reduced substantially (or even ceased completely) its activity during the last $10^5$ years \citep{lin09, kee12}. A second hypothesis proposed to explain the observed morphology of this object is that there was no decrease in the activity of the AGN in IC 2497, which is currently illuminating and heating Hanny's Voorwerp \citep{joz09, lin09, ram10}. In this scenario, the lack of X-ray emission from the AGN can be explained by the hypothesis that it is Compton-thick toward the observer, but not toward Hanny's Voorwerp. Based on the similarity between the scenarios proposed for the nuclear region of NGC 3621 and for Hanny's Voorwerp, we can say that the off-centered line-emitting blob we detected is possibly a miniature of Hanny's Voorwerp.

All the hypotheses described so far involve moderate or high obscuration toward the nucleus of NGC 3621. One could ask about the implications of such obscuration to the IR spectrum of NGC 3621. In order to discuss that, first of all, it is important to try to analyze the distribution of dust around the AGN in this galaxy. The $A_V$ values provided by the spectral synthesis in the region corresponding to the stellar nucleus were all smaller than 1.0 mag. In addition, as mentioned before, after the spectra in this area were extinction corrected, the resulting Balmer decrements (H$\alpha$/H$\beta$) were consistent with the expected value for Case B recombination, with a temperature of $10^4$ K and an electron density of $10^2$ cm$^{-3}$. In the hypothesis of a larger amount of dust, it would certainly not be located at the NLR or at the outer part of the nuclear stellar cluster, but actually much closer to the AGN. A distribution of dust in the form of a torus, for example, would probably not result in the observed scenario, where essentially no optical emission is detected close or at the nucleus. We believe that a distribution with a larger covering factor than that of a torus is required to produce this effect. In principle, if we could determine a representative distance range from the AGN and, as a consequence, a representative temperature range for most of the dust in the nuclear region of this object, we could estimate a wavelength range for the maximum of the IR emission. One possible approach is to take the sublimation radius of the dust as a minimum distance for the location of most of the dust and half of the FWHM of the PSF as a maximum distance for the dust. The problem of these assumptions is that they result in a large distance range, which also implies in a temperature range too large to provide valuable information about the IR emission maximum. Therefore, without more details about the distribution of dust (in particular, the minimum and maximum distances for the dust and also the geometry of the distribution), we cannot give much information about the IR implications of the obscured nucleus in this galaxy. The fact that this galaxy has a considerable AGN related IR emission was already confirmed by \citet{sat07}, using \textit{Spitzer} data. However, in order to determine properties like the temperature of the dust or the geometry of the distribution of dust, higher spatial resolution data, together with a larger spectral coverage, are required.

Although we believe that the scenarios presented above are the most likely to explain the observations of NGC 3621, a third scenario should also be mentioned here. It involves an off-centered AGN in the nuclear region of NGC 3621. It is well known that galaxy mergers may result in two SMBHs in the nuclear region of the merged remnant. The merging of the two black holes until their final coalescence goes through three phases. First, the black holes sink toward the center of the galaxy via dynamical friction and form a binary. After that, the binary is gravitationally bound and continues to decay, while its binding energy increases. Finally, the binary's separation decreases to the point where the emission of gravitational waves carries away the remaining angular momentum, allowing the black holes to coalesce (for more details, see Begelman et al. 1980, Merritt \& Milosavljevic 2005, and Merritt 2006). The asymmetric emission of gravitational waves carries linear momentum and, as a consequence of conservation of momentum, the resulting SMBH recoils (see Sundararajan et al. 2010 and references therein). An AGN ejected from the center of a galaxy usually carries the accretion disk and the broad-line region with it. Observational consequences of that are a spatial offset between the AGN and the nucleus of the galaxy and/or a spectral offset between the broad and narrow emission lines in the spectrum \citep{loe07, ble08, kom08a}. So far, a set of candidates for recoiling SMBHs have been found due to the detection of spatial offsets \citep{bat10, jon10, kos14, len14, men14b} or kinematic offsets \citep{kom08b, shi09a, shi09b, rob10}. On the other hand, only one candidate with both spatial and kinematic offsets (CXOC J100043.1+020637, also known as CID-42) has been found \citep{civ10, civ12, ble13}. We propose that the off-centered line-emitting blob in NGC 3621 could be a candidate for a recoiling SMBH, with an obvious spatial offset. Since there is no broad emission line in the spectrum of the off-centered emitting region, we could not detect any kinematic offset between broad and narrow lines. However, as mentioned before, the gas radial velocity (obtained from the [N II] $\lambda 6584$ emission line) and the stellar radial velocity measured in the area corresponding to the emitting region are not compatible with each other. If we assume that the kinematics of the gas emitting the narrow lines is being affected by the acreeting SMBH (which is a plausible hypothesis, considering that this galaxy has a significant amount of gas in its central region), we can say that the difference between the stellar radial velocity and the gas radial velocity is consistent with the recoiling SMBH scenario. 

The main problem of the hypothesis involving a recoiling SBMH is that, in this scenario, there is no AGN at the position of the stellar nucleus, while the X-ray data analyzed by \citet{gli09} seem to indicate otherwise. This hypothesis also requires the occurence of a merger. The two episodes of star formation revealed by the spectral synthesis are consistent with a merger, some $(5 - 9) \times 10^8$ years ago. However, an SA(s)d galaxy like NGC 3621 is not an expected outcome of a merger. Considering all of that, we conclude that the recoiling SMBH scenario is very unlikely in this object. Nevertheless, we cannot exclude the possibility that the recoil of the SMBH in NGC 3621 was caused by a merger between the nuclear SMBH and an intermediate-mass black hole. Such scenario would probably not alter the general morphology of the galaxy.

One final hypothesis that should be mentioned here is that the X-ray source in the nucleus of NGC 3621 could actually be an X-ray binary. X-ray binaries have hard X-ray spectra and luminosities, in the 2 - 10 keV spectral band, in the range of $10^{36} - 10^{39}$ erg s$^{-1}$ (e.g. Lewin et al. 1995). Since the X-ray luminosities of the central source detected by \citet{gli09} are $L_{2-10 keV}(low) = 5 \times 10^{37}$ erg s$^{-1}$ and $L_{2-10 keV}(high) = 2 \times 10^{39}$ erg s$^{-1}$, in the moderate and heavy absorption regimes, respectively, we cannot rule out the possibility that this source is actually an X-ray binary and not an AGN. This hypothesis is compatible with all the scenarios described above.

\section{Summary and conclusions}

We analyzed the morphology of the line-emitting areas in the nuclear region of NGC 3621, using an optical data cube obtained with GMOS-IFU. We verified that most of the line emission comes from a blob, located at a projected distance of $2\arcsec\!\!.14 \pm 0\arcsec\!\!.08$ (which is equivalent to $70.1 \pm 2.6$ pc) from the stellar nucleus, at PA $= 163\degr \pm 4\degr$. We believe that the line emission detected by \citet{sat07} and by \citet{bar09} was originated in this blob, as only diffuse emission was detected along the rest of the FOV, with a deficit of emission at the position of the nucleus. Diagnostic diagram analysis revealed that the emission-line ratios of the spectrum of the emitting-blob are compatible with those of a Seyfert 2 galaxy. The nuclear diffuse emission is also characteristic of a Seyfert 2, but with a lower ionization degree, which reveals an AGN emission with a lower ionization parameter. The nuclear spectrum is also probably contaminated by the emission from H II regions.

The line-emitting blob does not show distinct spectral stellar properties in comparison to the surroundings. The gas radial velocity (obtained from the [N II] $\lambda 6584$ emission line) at the blob ($V_{gas}(blob) = -46 \pm 5$ km s$^{-1}$), relative to the nucleus, is not compatible, even at the 3$\sigma$ level, with the stellar radial velocity ($V_*(blob) = -10 \pm 3$ km s$^{-1}$), at the same area. This indicates that the gas kinematics in the emitting blob is apparently disconnected from the stellar kinematics there. An HST-WFC3 image of the nuclear region of NGC 3621, in the \textit{U} band, does not show traces of the blob or hot stars to ionize the gas in this area.

The first scenario we propose to explain the unusual morphology of the emitting areas in the nuclear region of NGC 3621 is that the line-emitting blob is a ``fossil'' emission-line region or a light ``echo'' from the previously brighter AGN in the galaxy. A simulation performed with the Cloudy software, assuming a central AGN, provides emission-line ratios for the blob compatible with the observed values, at 1$\sigma$ or 2$\sigma$ levels. This simulation also provides a bolometric luminosity of $L_{bol}(simulated) = 4 \times 10^{41}$ erg s$^{-1}$ for the AGN. Considering the current estimates of the AGN bolometric luminosity, obtained assuming different column densities toward the nucleus (moderate and heavy absorption regimes; Gliozzi et al. 2009), we conclude that the AGN must be now $\sim 13 - 500$ times fainter than it was in the past. The decrease of the AGN bolometric luminosity must have occurred during the last $\sim 230$ years. The detection of an X-ray source at the position of the stellar nucleus of NGC 3621 \citep{gli09} is consistent with the AGN shutdown scenario. A large obscuration of the central source could explain the fact that only diffuse optical emission is detected at the position of the X-ray source. 

A second scenario to explain the observed properties of the nuclear region of NGC 3621  involves no decrease of the AGN bolometric luminosity. In that case, the AGN is highly obscured toward the observer but not toward the line-emitting blob. 

The third scenario considered here establishes that the line-emitting blob is a recoiling SMBH, after the coalescence of two black holes. The fact that the stellar radial velocity and the gas radial velocity, at the position of the blob, are not compatible with each other is consistent with the recoiling SMBH hypothesis. The possible merger revealed by the two episodes of star formation, detected by the spectral synthesis, is actually required in this scenario. However, the nuclear X-ray source detected by \citet{gli09} suggests the presence of an AGN at the stellar nucleus, which is not consistent with the recoiling SMBH hypothesis. In addition, an SA(s)d galaxy like NGC 3621 is the least likely outcome of a merger. Therefore, we consider this scenario the most unlikely to explain the morphology of the emitting areas in the nuclear region of NGC 3621. Nevertheless, we cannot discard the possibility that a recoiling SMBH here could be the result of a merger between the nuclear SMBH and an intermediate-mass black hole, which would probably not affect the morphology of the galaxy.

One final hypothesis, compatible with all previous scenarios, is that the central X-ray source detected by \citet{gli09} is not an AGN, but an X-ray binary. The luminosity of the source ($L_{2-10 keV}(low) = 5 \times 10^{37}$ erg s$^{-1}$ and $L_{2-10 keV}(high) = 2 \times 10^{39}$ erg s$^{-1}$, in the moderate and heavy absorption regimes, respectively) is consistent with this hypothesis, which should be taken into account in future studies of NGC 3621.

\acknowledgments

Based on observations obtained at the Gemini Observatory (processed using the Gemini IRAF package), which is operated by the Association of Universities for Research in Astronomy, Inc., under a cooperative agreement with the NSF on behalf of the Gemini partnership: the National Science Foundation (United States), the National Research Council (Canada), CONICYT (Chile), the Australian Research Council (Australia), Minist\'{e}rio da Ci\^{e}ncia, Tecnologia e Inova\c{c}\~{a}o (Brazil) and Ministerio de Ciencia, Tecnolog\'{i}a e Innovaci\'{o}n Productiva (Argentina). We thank FAPESP for support under grants 2012/02268-8 (RBM) and 2011/51680-6 (JES) and also an anonymous referee for valuable comments about this article.

{\it Facilities:} \facility{Gemini:South(GMOS)}.

\end{document}